\documentstyle[psfig]{l-aa}        

\def\ros{{\sl ROSAT }}

\def\heao{{\sl HEAO 1 }}

\def\G{$\Gamma_{\rm x}$ } 
\def\NH{$N_{\rm H}$}

\def\degs{\ifmmode ^{\circ}\else$^{\circ}$\fi}
\def\arcmin{\ifmmode ^{\prime}\else$^{\prime}$\fi}
\def\arcsec{\ifmmode ^{\prime\prime}\else$^{\prime\prime}$\fi}

\def\approxlt{\mathrel{\hbox{\rlap{\lower.55ex \hbox {$\sim$}}
        \kern-.3em \raise.4ex \hbox{$<$}}}}
\def\approxgt{\mathrel{\hbox{\rlap{\lower.55ex \hbox {$\sim$}}
        \kern-.3em \raise.4ex \hbox{$>$}}}}

\begin{document}
 
   \thesaurus{03         
              (11.01.2;  
               11.09.1;  
               11.17.2;  
               11.19.1;  
               13.25.2)  
}

   \title{Soft X-ray properties of the Seyfert\,1.8 galaxy NGC 3786} 
   \author{Stefanie Komossa, Henner Fink$^{\dagger}$} 

   \offprints{St. Komossa, skomossa@mpe-garching.mpg.de \\
       $\dagger$ deceased in Dec. 1996}
 
  \institute{Max-Planck-Institut f\"ur extraterrestrische Physik,
             85740 Garching, Germany\\
       }

   \date{Received 3 February 1997; accepted June 1997}

   \maketitle\markboth{St. Komossa, H. Fink~~~Soft X-ray properties of NGC 3786}{}  

   \begin{abstract}
An analysis of survey and  pointed \ros PSPC observations of the
Seyfert 1.8 galaxy NGC 3786 has revealed  
interesting spectral and temporal behaviour.   
The spectrum is found to show clear signs of excess absorption, 
and there is evidence that part (or all) of it
is caused by the presence of a warm absorber. 
The soft X-ray spectral properties are discussed in 
the context of the Sy 1.8 classification 
of NGC 3786, and are combined with published optical data
in an effort to     
discriminate between several possible absorption models.
The one involving a warm absorber {\em with internal dust}, located 
between BLR and NLR, 
is favoured and shown to account successfully for  
the high observed broad line reddening   
as well as for the strong X-ray variability (a factor $\sim$10) we detect 
between the 1990 and 1992 observation.   
 
      \keywords{Galaxies: active -- individual: NGC 3786 -- emission lines --
Seyfert -- X-rays: galaxies 
               }

   \end{abstract}
 
\section{Introduction}

NGC 3786 (Mrk 744) is a spiral galaxy 
at a redshift of $z$=0.009 that hosts an active nucleus.
The galaxy forms a pair with its peculiar companion
NGC 3788 and is included in several studies of Seyferts with companions
(e.g. Keel et al. 1985, Keel 1996; Rafanelli et al. 1995). 

The Seyfert nature of NGC 3786 was first noted by Afanas'ev et al. (1979a).
Detailed optical spectroscopy was presented by Afanas'ev et al. (1979b)
and by Goodrich \& Osterbrock (1983; GO83 hereafter). GO83
classified the spectrum as Seyfert 1.8 according to the scheme of
Osterbrock (1981). 
The galaxy is 
included in several investigations of the nature of Sy 1.8 and Sy 1.9 galaxies
(e.g. Goodrich 1989, 1990; De Zotti \& Gaskell 1985; Maiolino \& Rieke 1995). 
One of the characteristics of the intermediate type Seyferts 
is steep Balmer decrements of the broad lines (Osterbrock 1981). 
The cause for this is still under discussion, 
as is their place in the unified model of Seyfert galaxies 
(e.g. Lawrence 1987, Antonucci 1993). One interpretation is that intermediate type
Seyferts are viewed
through dusty material which causes the high observed Balmer line
flux ratios by 
extinction, and which may be identified with an inner (e.g. Goodrich 1995)
or outer (Maiolino \& Rieke 1995) torus.   
   
Recently, Nelson (1996) presented strong evidence for a thermal dust origin 
of the IR continuum of NGC 3786, through the analysis of a correlated, but time-delayed 
optical-IR outburst. 

In X-rays, an upper limit for the count rate in the \heao survey is 
reported by Persic et al. (1989).   
We are not aware of any more detailed
X-ray study of NGC 3786.  
For the present investigation, we have analyzed the survey and archival pointed \ros 
(Tr\"umper 1983) observations of NGC 3786  
performed with the PSPC
(Pfeffermann et al. 1987).   
Besides other soft X-ray properties, the amount of X-ray absorption
present is particularly interesting in view of the Sy 1.8 character of NGC 3786,
and \ros with its (0.1--2.4 keV) energy range is well suited for such a study. 
 
The paper is organized as follows: 
In Sect. 2 we present the observations. 
The analysis of the data with respect to their spectral and temporal properties 
is described in  
Sects. 3 and 4, respectively. 
In Sect. 5,  
the X-ray spectrum is discussed in light of the
Sy 1.8 classification of NGC 3786, and the inferences drawn 
from the X-ray data are
combined with published optical observations.  
A summary and the conclusions are given in Sect. 6. 

A distance of 54 Mpc is adopted for NGC 3786 assuming a Hubble constant of $H_{\rm o}$ = 50 km/s/Mpc
and the galaxy to follow the Hubble flow. 
If not stated otherwise, cgs units are used throughout.

\section{Data reduction}

\subsection{Pointed data} 
The observation was performed with the \ros PSPC on Dec. 6, 1992, 
with NGC 3786 in the centre of the field of view.   
The effective exposure time was 2.9 ksec.   
In total, 17~X-ray sources were detected with a likelihood $\ge$ 10 within the field of view.
The positions of those sources in the vicinity of NGC 3786 are shown in Fig. 1, overlaid 
on an optical image from the digitized Palomar sky survey (E plate). 
The X-ray position of the pointlike central source is at 
$\alpha = 11^{\rm h}39^{\rm m}42.7^{\rm s}$, $\delta$ = 31\degs54\arcmin39.0\arcsec~ (J\,2000), 
consistent with the optical 
position $\alpha = 11^{\rm h}39^{\rm m}42.6^{\rm s}$, $\delta$ = 
31\degs54\arcmin33.4\arcsec~ (Clements 1983).   
The uncertainty in the X-ray position 
resulting from the telescope boresight error is of the order of 10\arcsec~ -- 15\arcsec~
(Briel et al. 1994).  
 
For further analysis, source photons were extracted
within a circle centered around the X-ray position of NGC 3786. 
The background was determined from the inner 19\arcmin~ of the field of view, 
after the removal of all detected sources.   
Vignetting
and dead-time corrections were applied to the data 
using the EXSAS software package (Zimmermann et al. 1994).
The mean source count rate is 0.36 $\pm {0.01}$ cts/s.
The other X-ray sources shown in Fig. 1 are weak, with count rates of about
0.01, 0.01, and 0.007 cts/s, respectively, ordered by increasing distance from the target source.
For the spectral analysis source photons in
the amplitude channels 11-240 were binned 
according to a constant signal/noise ratio of 7$\sigma$. 
For the temporal analysis the minimal bin size in time was 400 s to account for the
wobble mode of the observation. 
%
\begin{figure}[thbp]
 \hspace{0.6cm} 
      \vbox{\psfig{figure=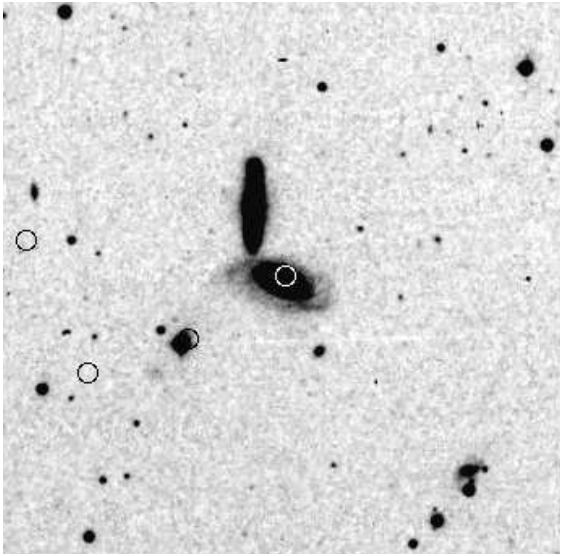,width=7.4cm,%
          bbllx=5.7cm,bblly=13.2cm,bburx=16.1cm,bbury=23.5cm,clip=}}\par
 \caption[N3786_ima]{X-ray sources detected 
in a 10\arcmin~ $\times$ 10\arcmin~ field 
around NGC 3786 superimposed on an optical image. The circles drawn around
the X-ray source positions are of 10\arcsec~ radius.  
   }
 \label{N3786_ima}
\end{figure}
%
\subsection{Survey data}
The sky field around NGC 3786 was observed during the \ros all-sky survey (RASS)
from Nov. 24 -- 26, 1990, with an effective exposure time of about 400 s.
Emission from NGC 3786 is detected, but found to be much weaker than in 
the pointed observation.
After determining the background from a source-free region along
the scanning direction of the telescope
and correcting the data for vignetting,
we find a source count rate
of 0.04 $\pm{0.01}$ cts/s. Due to the low number of detected photons 
no spectral analysis of the RASS observation can be performed and 
the following results (Sect. 3) refer to the pointed data only. 
The detected X-ray variability
will be further discussed in Sect. 4.  

\section{Spectral analysis} 

\subsection{Standard spectral models}
\begin{figure}[t]      
      \vbox{\psfig{figure=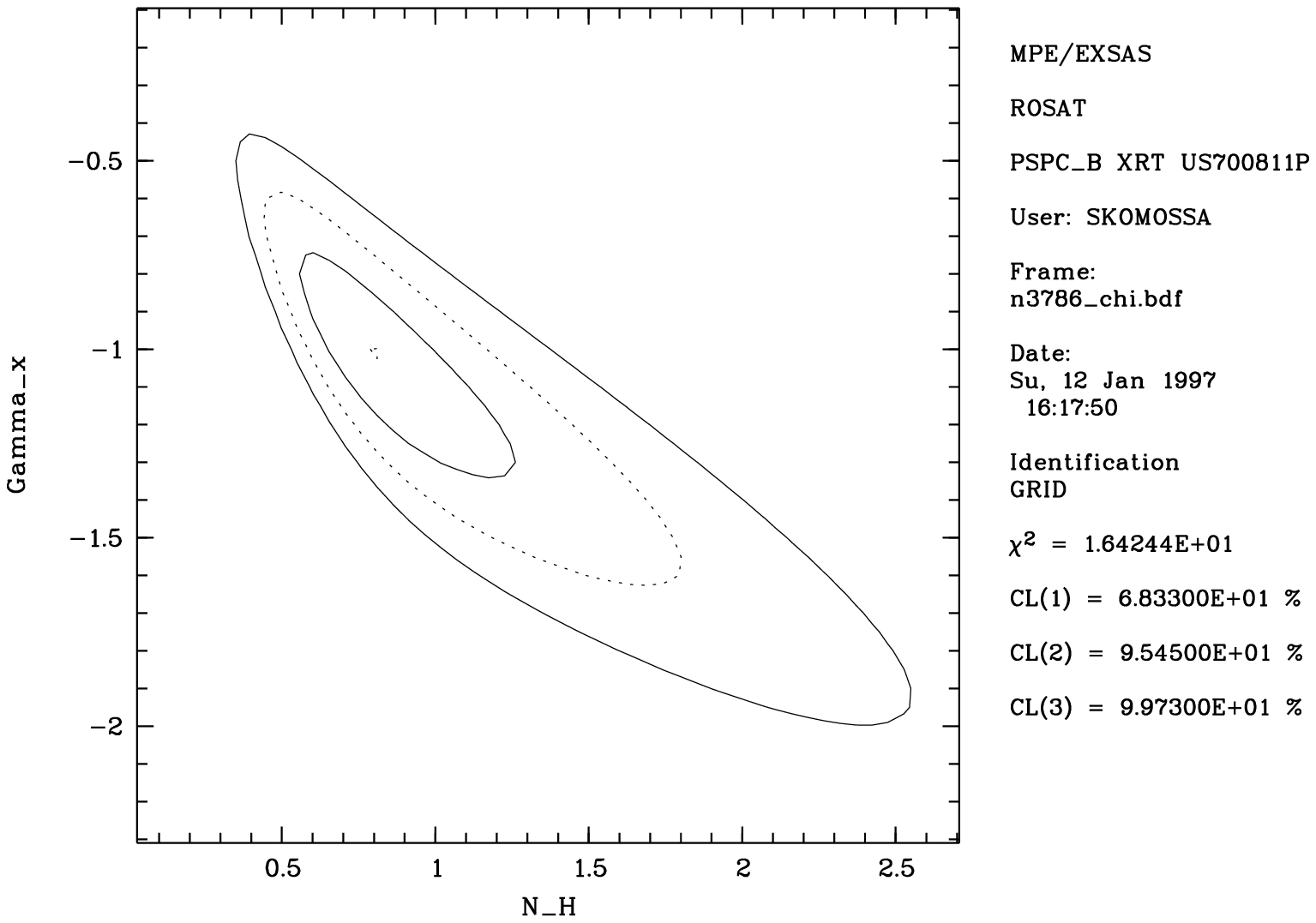,width=7.3cm,%
          bbllx=2.2cm,bblly=1.1cm,bburx=14.5cm,bbury=12.2cm,clip=}}\par
      \vspace{-0.4cm}
\begin{picture}(12,25) 
\put(83,129){+}
\end{picture}
\vspace{-1.0cm}
 \caption[chi]{Error ellipses in \G, \NH~(in units of 10$^{21}$ cm$^{-2}$)
 for the single-powerlaw description of the
X-ray spectrum. The two dimensional contours are shown for confidence levels
of 68.3, 95.5 and 99.7\%. The Galactic absorbing column density
is \NH~ = 0.222 $\times$ 10$^{21}$ cm$^{-2}$.
}
 \label{chi}
\end{figure}

A single powerlaw with cold absorption column as a free parameter
 acceptably fits the X-ray spectrum ($\chi^2_{\rm red}$ = 1.0), 
although some systematic residuals remain around 0.6--1.0 keV. The deduced 
power law is unusually flat, with a photon index $\Gamma_{\rm{x}} \simeq$ --1.0. 
An absorbing column of \NH~ $\simeq 8~\times 10^{20}$ cm$^{-2}$ is
found, exceeding the     
Galactic value towards
NGC 3786 of \NH~ = 2.22 $\times$ 10$^{20}$ cm$^{-2}$ (Murphy et al. 1996).   
Fig. \ref{chi} displays the error ellipses for this model description. 

Among various non-powerlaw spectral models compared to the 
data, 
including thermal bremsstrahlung, a Raymond-Smith model with cosmic abundances
(Raymond \& Smith 1977), and emission from an accretion disk after Shakura \& Sunyaev (1973),
only a black body gives an acceptable fit with 
$kT_{\rm bb} \simeq$ 0.5 keV ($\chi^2_{\rm red}$ = 1.2). 
The cold column density is found to be consistent with the Galactic value
in this case.    
Again, some systematic residuals remain around 0.6--1.0 keV.  

The application of a powerlaw model with an additional absorption 
edge gives an excellent fit ($\chi^2_{\rm red}$ = 0.7). 
We find an edge energy of $E \approx$ 0.8 keV   
and a photon index of \G=--1.9$\pm 0.3$. The cold absorption,
with \NH~= (1.7$\pm 0.3$) $\times 10^{21}$ cm$^{-2}$, largely 
exceeds the Galactic value.   
The edge energy is near that expected for OVII ($E$ = 0.74 keV) and OVIII
($E$ = 0.87 keV). 
This is very suggestive of the presence of a warm absorber, 
which is discussed in more detail in the next section. 

  \begin{figure}[thbp]
      \vbox{\psfig{figure=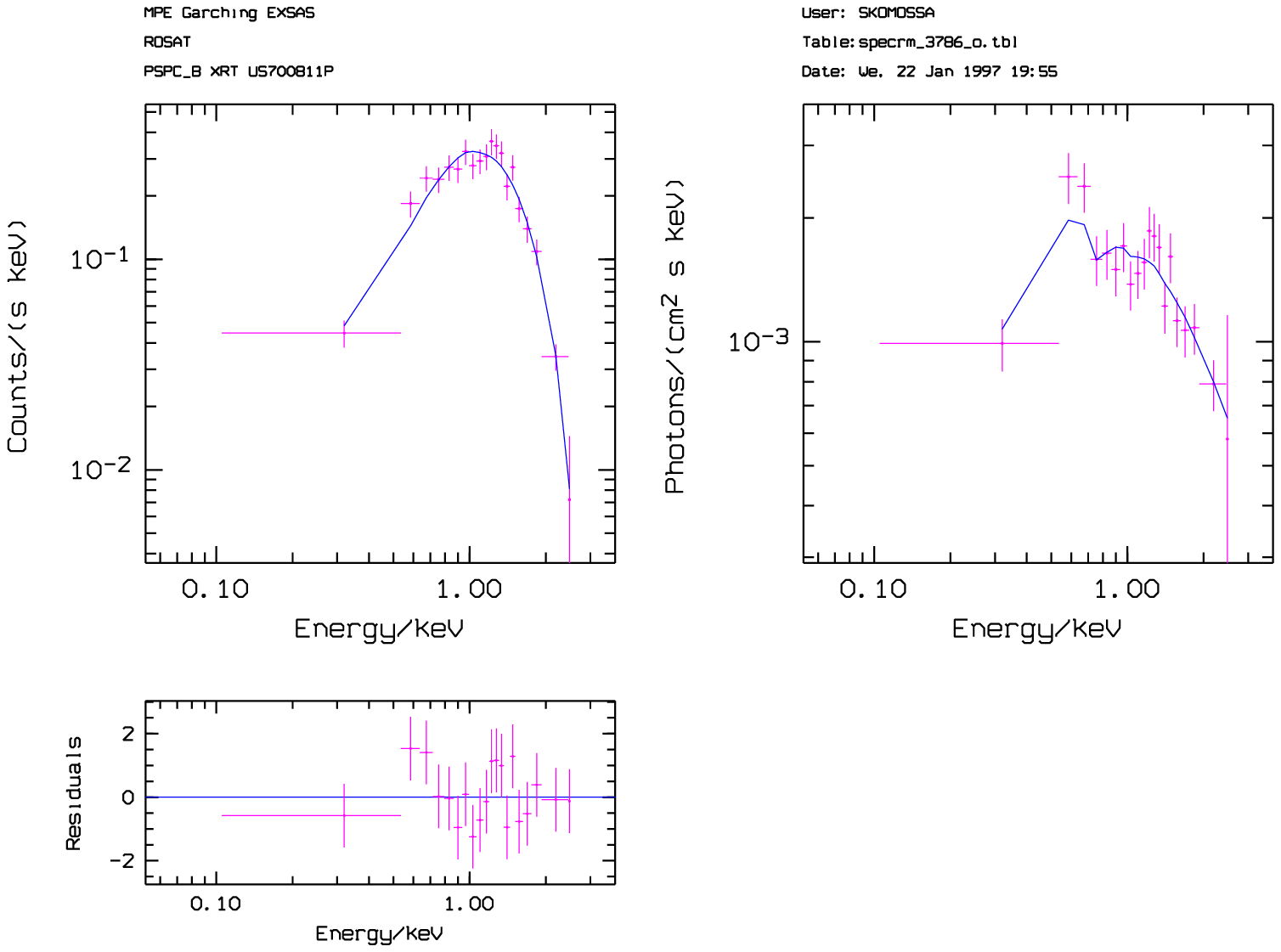,width=6.85cm,%
          bbllx=2.5cm,bblly=1.1cm,bburx=10.1cm,bbury=11.7cm,clip=}}\par
            \vspace{-0.5cm}
      \vbox{\psfig{figure=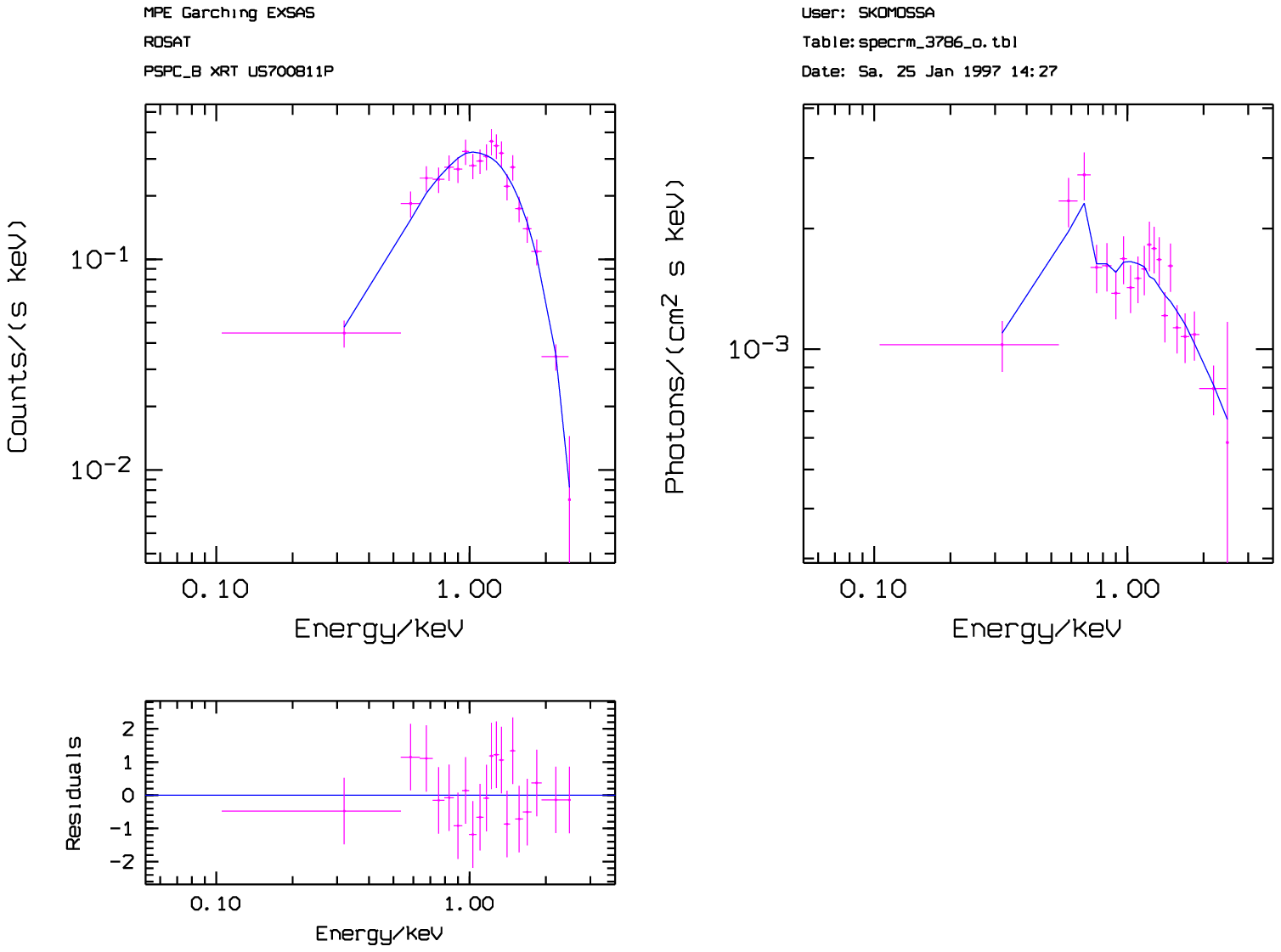,width=6.85cm,%
          bbllx=2.5cm,bblly=1.1cm,bburx=10.1cm,bbury=4.5cm,clip=}}\par
            \vspace{-0.5cm}
      \vbox{\psfig{figure=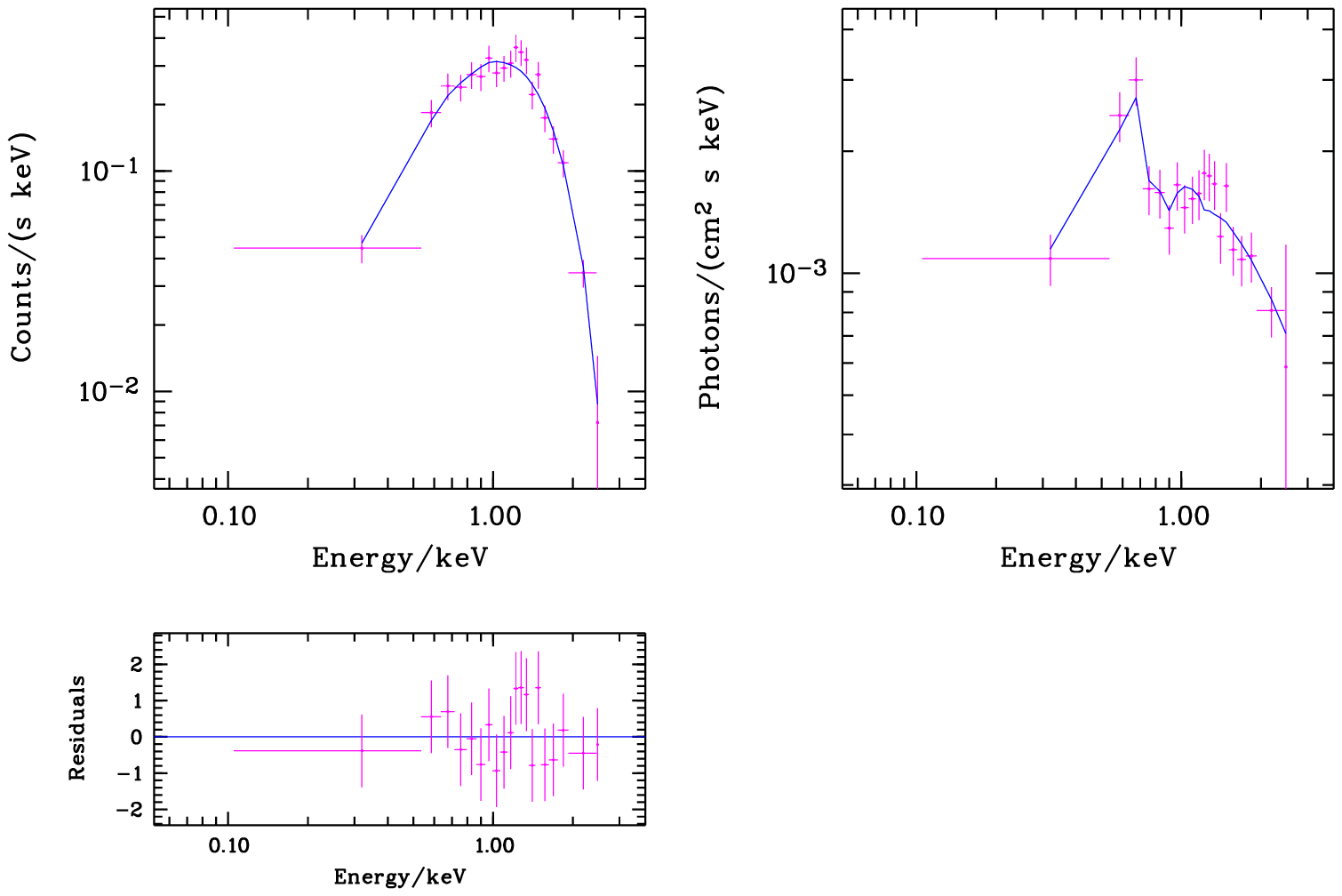,width=6.85cm,%
          bbllx=2.5cm,bblly=1.1cm,bburx=10.1cm,bbury=4.5cm,clip=}}\par 
            \vspace{-0.5cm}
      \vbox{\psfig{figure=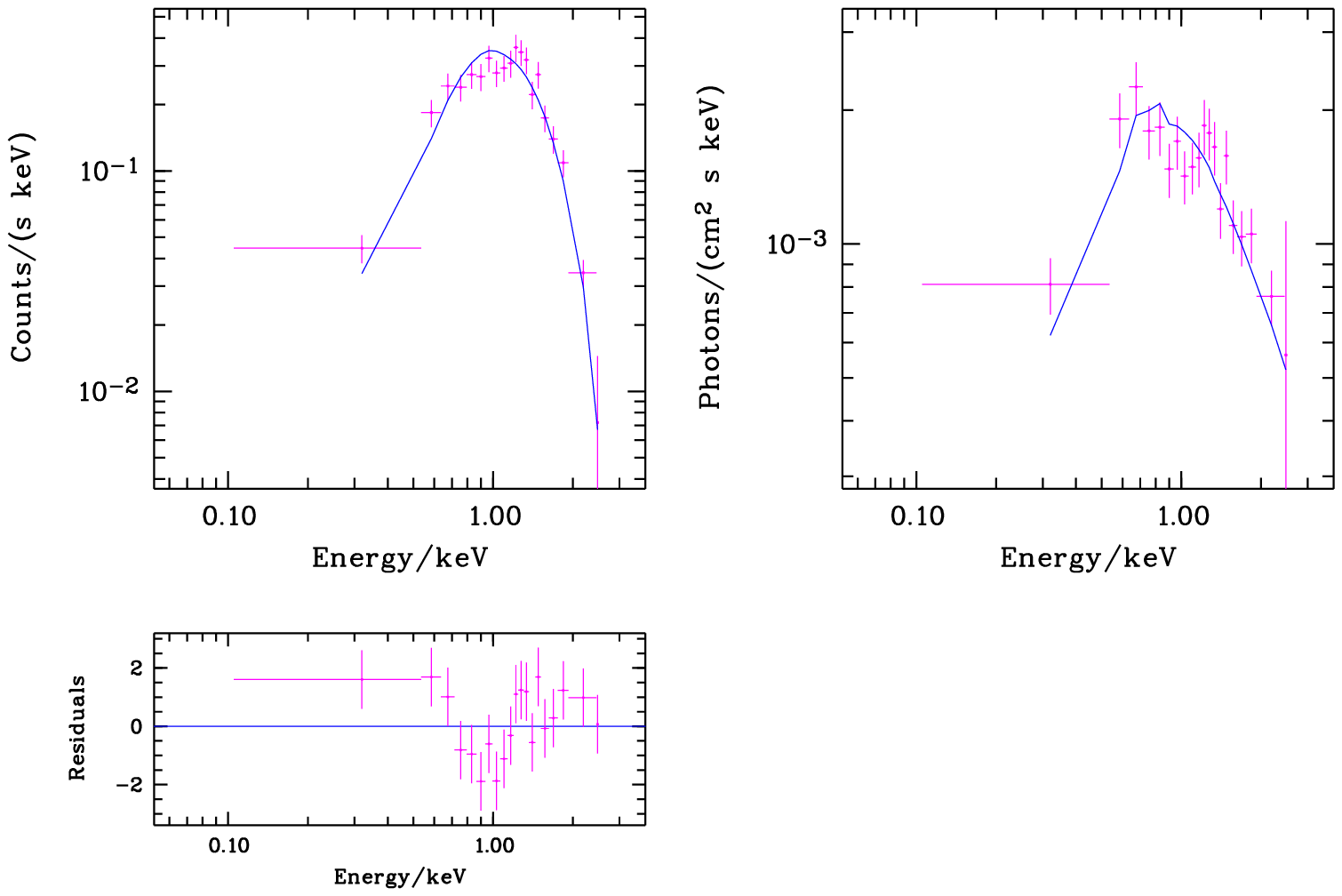,width=6.85cm,%
          bbllx=2.5cm,bblly=1.1cm,bburx=10.1cm,bbury=4.5cm,clip=}}\par
            \vspace{-0.5cm}
      \vbox{\psfig{figure=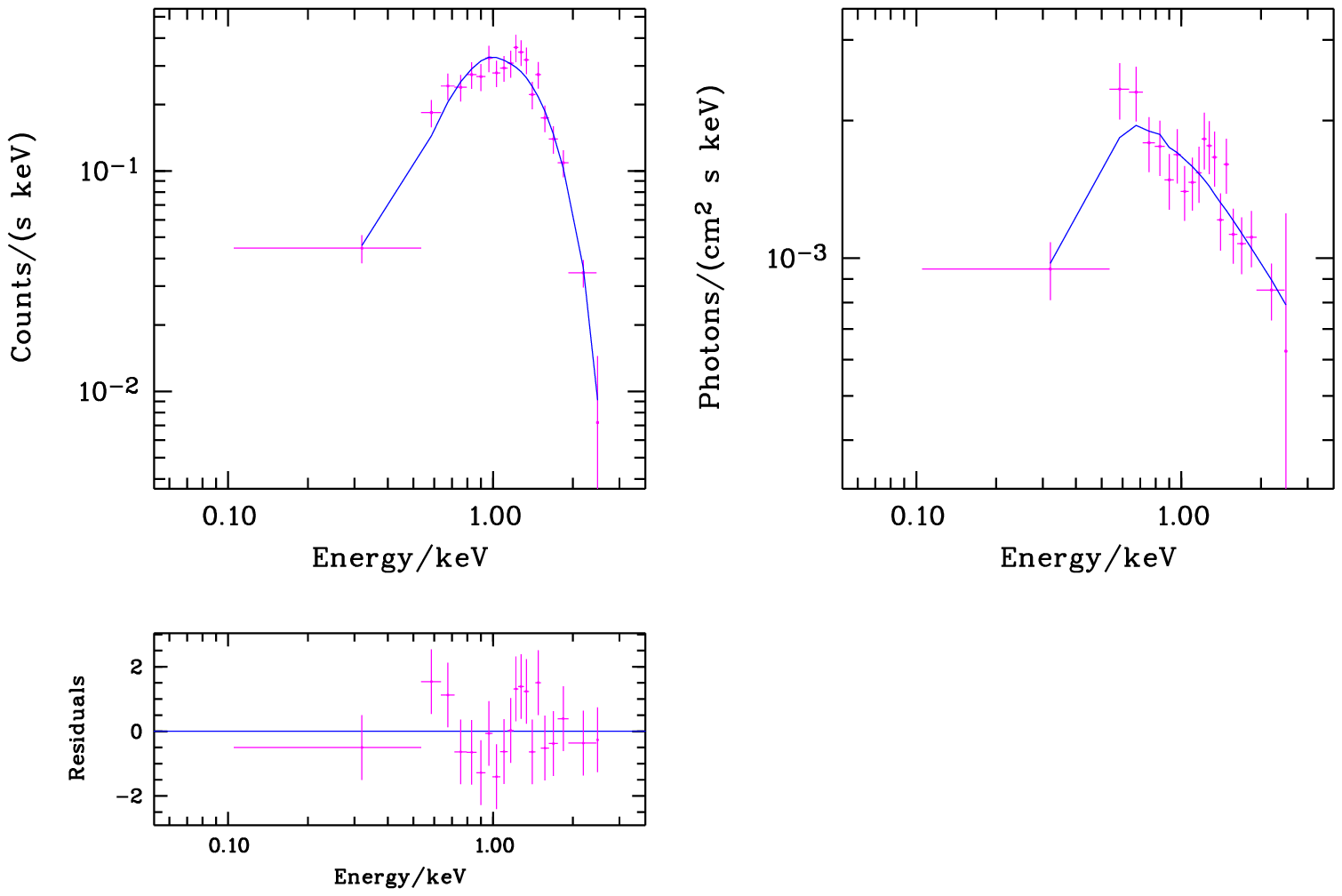,width=6.85cm,%
          bbllx=2.5cm,bblly=1.7cm,bburx=10.1cm,bbury=4.5cm,clip=}}\par
            \vspace{-0.5cm}
\begin{center}
\begin{picture}(12,25)\thicklines
\put(-60,540){(a)}     
\put(-60,290){(b)} 
\put(-60,216){(c)}
\put(-60,113){(d)}
\put(-60,70){(e)}   
\end{picture}
\end{center}
\vspace{-1.6cm} 
\caption[SEDx]{The upper panel (a) shows the observed X-ray spectrum of NGC 3786 (crosses)
and the best-fit dusty warm absorber model with fixed \NH~ (solid line; model 6 of Table 1). 
The second panel displays
the fit residuals for this model. In the third panel (b) the residuals from the 
same model, but with \NH~ as additional free parameter (model 7), are depicted
and in the fourth panel (c) those of the description involving a dust-free warm absorber
(model 5). 
For comparison, the residuals resulting from
a single powerlaw fit to the data  
are shown in the fifth panel~(d).  
In all models so far, 
the powerlaw index was fixed to \G = --1.9.  
The lowest panel (e) displays the residuals for the powerlaw fit with free 
photon index (model 1;  
note the slightly different scales in the ordinate).
  }
\label{SEDx}
\end{figure}
\subsection{Warm absorber models}

\subsubsection{Model properties and assumptions}

To calculate the X-ray spectral absorption structure resulting from a warm absorber 
(WA) we used the 
photoionization code {\em Cloudy} (Ferland 1993).   
The warm material was assumed to be photoionized by the continuum emission of the 
nucleus, to be one-component and of constant density. 
Solar abundances (Grevesse \& Anders 1989) were adopted. 
In those models that include dust mixed with the warm gas,
the dust composition and grain size distribution 
were like those of the Galactic diffuse interstellar medium, 
and the chemical abundances depleted correspondingly,  
as in Ferland (1993).

The spectral energy distribution incident on the warm material  
was chosen to be a `mean Seyfert' continuum. It consists of 
an UV-EUV powerlaw of energy index $\alpha_{\rm uv-x}$=--1.4 
extending up to 0.1 keV, a mean optical to radio continuum after 
Padovani \& Rafanelli (1988), a break at 10$\mu$m and an index 
$\alpha$ = --2.5 $\lambda$-longwards. For comments on the
usually weak influence of non - X-ray spectral parts on the warm absorption  
structure see Komossa \& Fink (1997a). 
 
Those two properties characterizing the warm absorber 
that can be directly extracted from X-ray spectral fitting 
are the hydrogen column density $N_{\rm w}$ of the ionized material and the     
ionization parameter $U$. The latter is defined as  
\begin {equation}
U=Q/(4\pi{r}^{2}n_{\rm H}c)
\end {equation} 
where $Q$ is the
number rate of incident photons above 13.6 eV, $r$ is the distance between
nucleus and warm absorber, $n_{\rm H}$ is the hydrogen density 
(fixed to 10$^{9.5}$ cm$^{-3}$ unless noted otherwise) 
and $c$ the speed of light.  

   \begin{table*}     
     \vspace{-0.5cm}
     \caption{X-ray spectral fits to NGC 3786 (pl = powerlaw, bb = black body, 
                  wa = warm absorber). 
        The errors are quoted at the 68\% confidence level. 
        The (0.1--2.4 keV) luminosity corrected for cold (and warm) absorption
        is $L_{\rm x}$ = (2.6, 1.9, 6.7)$\times 10^{42}$ erg/s for models 1, 3, 4, respectively. 
        }
     \label{fitres}
      \begin{tabular}{llccclcc}
      \hline
      \noalign{\smallskip}
        model & $N_{\rm H}$ & log $U$ & log $N_{\rm w}$ & Norm$_{\rm pl}$$^{(2)}$
                            & ~$\Gamma_{\rm x}$ or $kT_{\rm bb}$ & Norm$_{\rm bb}$   
                            & $\chi^2(d.o.f)$ \\
       \noalign{\smallskip}
      \noalign{\smallskip}
           & [10$^{21}$ cm$^{-2}$] & & [cm$^{-2}$] & [ph/cm$^2$/s/keV] & ~~~~~~~~~[keV] & [ph/cm$^2$/s]  
                  & \\
       \noalign{\smallskip}
      \hline
      \hline
      \noalign{\smallskip}
1~~~~ pl & 0.8$\pm{0.4}$ & - & - & (2.0$\pm{0.7}$)~10$^{-3}$ & --1.0$\pm{0.2}$ & - & 16.4(17) \\ 
      \noalign{\smallskip}
      \hline
      \noalign{\smallskip}
2a~~~ pl & 0.222$^{(1)}$ & - & - & (1.5$\pm{0.1}$)~10$^{-3}$ & --0.3$\pm{0.1}$ & - & 45.0(18) \\ 
      \noalign{\smallskip}
      \hline
      \noalign{\smallskip}
2b~~~ pl & 2.2$\pm{0.4}$ & - & - & (3.1$\pm{0.3}$)~10$^{-3}$ & --1.9$^{(3)}$ & - & 26.7(18) \\  
      \noalign{\smallskip}
      \hline
      \noalign{\smallskip}
3~~~~ bb & 0.23$\pm{0.1}$ & - & - & - & 0.49$\pm{0.06}$ & (3.4$\pm{0.3}$)~10$^{-3}$ &  20.4(17) \\ 
      \noalign{\smallskip}
      \hline
      \noalign{\smallskip}
4~~~~ wa & 0.222$^{(1)}$ & --1.18$\pm{0.06}$ & 21.58$\pm{0.04}$ & (5.0$\pm{0.4}$)~10$^{-5}$
                & --1.9$^{(3)}$ & - & 14.3(17) \\
      \noalign{\smallskip}
      \hline
      \noalign{\smallskip}
5~~~~ wa & 1.8$\pm{0.4}$ & --0.31$\pm{0.22}$ & 21.78$\pm{0.16}$ & (5.9$\pm{1.1}$)~10$^{-5}$  
                & --1.9$^{(3)}$ & - & 10.9(16) \\
      \noalign{\smallskip}
      \hline
      \noalign{\smallskip}
6~~~~ dusty wa & 0.222$^{(1)}$ & --0.81$\pm{0.04}$ & 21.71$\pm{0.02}$ & (5.2$\pm{0.4}$)~10$^{-5}$  
                & --1.9$^{(3)}$ & - & 14.1(17) \\ 
      \noalign{\smallskip}
      \hline
      \noalign{\smallskip}
7~~~~ dusty wa & 1.2$\pm{0.4}$ & --0.30$\pm{0.25}$ & 21.74$\pm{0.23}$ & (5.4$\pm{0.9}$)~10$^{-5}$  
                & --1.9$^{(3)}$ & - & 12.3(16) \\
      \noalign{\smallskip}
      \hline
      \noalign{\smallskip}
  \end{tabular}

\noindent{\small $^{(1)}$ fixed to the Galactic value, $^{(2)}$ powerlaw flux, usually given at 1 keV, 
except for the models involving a warm absorber where it is given at 10 keV to represent the undistorted pl,
$^{(3)}$ fixed    
}
   \end{table*}

\subsubsection{Model results} 

For the comparison of the warm absorber model with the X-ray spectrum of
NGC 3786, \G was fixed to --1.9, since we particularly wanted to test
whether the data could be reconciled with a more typical Seyfert spectrum
of canonical photon index. This value of \G is also suggested by 
the single-edge fit to the spectrum.
The value of the cold absorbing column is less certain,
and may well exceed the Galactic value. 
In the following, we in turn discuss several absorption models. \\

(i) {\em Dust-free WA.}
In a first step, the value of the cold absorbing column 
was fixed to the Galactic value, $N_{\rm H}^{gal}$ = 2.22 $\times$ 10$^{20}$ cm$^{-2}$.    
This model provides a very good fit ($\chi^2_{\rm red}$ = 0.8)
with    
an ionization parameter of $\log U \simeq -1.2 $ and a column density 
of the ionized material of $\log N_{\rm w} \simeq 21.6$ 
(Table 1, model 4). 
The observed X-ray flux is $f$ = 0.55$\times 10^{-11}$ erg/cm$^2$/s, 
corresponding to an intrinsic (0.1--2.4 keV) luminosity corrected for cold and warm absorption
of $L_{\rm x} = 6.7 \times 10^{42}$ erg/s.
Since the density of the warm material is essentially
unconstrained by the X-ray spectral fits, only the density-scaled distance
of the material from the nucleus is known. Assuming $\log n_{\rm H}$ = 9.5
translates to a distance of $r \simeq$ 0.016 pc (using Eq. (1) and $Q$ from
integrating over the SED).  
The predicted warm-absorber intrinsic 
luminosity in H$\beta$ is $L_{\rm H\beta} \simeq 2 \times 10^{40}$ erg/s,
assuming full covering.     

Due to the possibility of partially compensating effects in the cold and warm column,  
\NH~ is not
well constrained in the spectral fits. 
Nevertheless, in a second step, it was left as a free parameter.
The best-fit column exceeds the Galactic value   
and a higher ionization parameter, $\log U \simeq -0.3$, 
is found for the warm absorber. 
Residuals around 0.6--1.0 keV  
are slightly better removed by this model than the previous one
(Fig. \ref{SEDx}c; Table 1, model 5). 

(ii) {\em Dusty WA.} 
Finally, we also applied a dusty warm absorber model to the data. 
Warm material with internal dust was 
found to reproduce successfully the X-ray spectrum and other
spectral properties of NGC 3227 (Komossa \& Fink 1996, 1997b). 
Dusty warm gas was first suggested to exist in 
the infrared loud quasar IRAS 13349+2438 (Brandt et al. 1996).  
To account for the modification of the X-ray absorption 
structure due to the presence of dust (Komossa \& Fink 1997a,b), 
we re-calculated the warm absorber models, 
now including dust with Galactic ISM
properties. 
Again, a successful fit is obtained.  
We find $\log U \simeq -0.8$ and
$\log N_{\rm w} \simeq 21.7$ for fixed cold Galactic column (Fig. \ref{SEDx}a;
Table 1, model 6), and 
$\log U \simeq -0.3$, $\log N_{\rm w} \simeq 21.7$, \NH $\simeq 1.2 \times 10^{21}$ cm$^{-2}$
for free cold column (Fig. \ref{SEDx}b; Table 1, model 7).   

A value of the gas density of log $n_{\rm H}$ = 8 was used
to calculate the models that include dust, to ensure dust survival. 
The density-scaled distance of the warm
material is given by $r \simeq$ 0.055/$\sqrt{n_{8}}$ pc, 
where $n_{\rm H} = 10^{8} n_{8}$ cm$^{-3}$.  
The emissivity of the gas in optical-UV emission lines 
is reduced as compared to the dust-free absorber. 

Below (Sect. 5) we will try to select among the X-ray absorption models the one 
that accounts best for 
the observed properties of NGC 3786  
in combining the X-ray spectral results with published multi-wavelength observations
of NGC 3786 and constraints from X-ray variability (Sect. 4). 
For the sake of brevity, the different spectral models 
will occasionally be referred to by their number as given in Table 1.

\section {Temporal analysis}

  \begin{figure}[thbp]
      \vbox{\psfig{figure=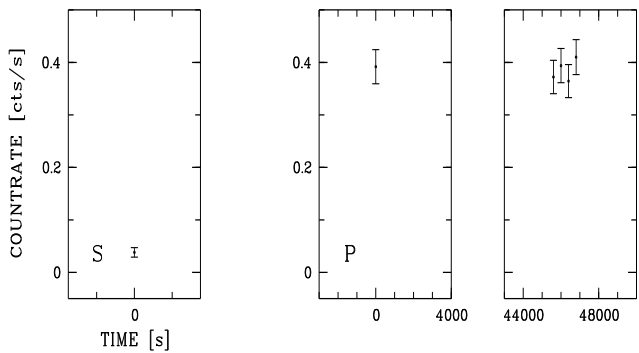,width=8.9cm,height=5.3cm,%
          bbllx=2.85cm,bblly=5cm,bburx=16.5cm,bbury=11.3cm,clip=}}\par
      \vspace{-0.6cm}
 \caption[light]{X-ray lightcurve of NGC 3786, binned to time intervals of 400 s. The time
is measured in seconds from the start of the observation; `S' denotes survey, `P' pointing.
   }
 \label{light}
\end{figure}
The X-ray lightcurve of NGC 3786 is shown in Fig. \ref{light}.
Due to the wobble in the telescope's pointing direction which
causes a time dependent shadowing of the source's image by
the supporting grids,  
a reliable estimate of the source count rate is 
possible only by averaging the count rate over an entire wobble period
of 400s. 
Since, usually, the total exposure time per satellite's orbit is not
an integer multiple of 400s, incomplete last bins were rejected
in producing the X-ray lightcurve.   
 
The count rate during the pointed observation is constant within the 
errors. The source is found to be strongly variable on a longer timescale,
however, since it is much weaker during the  
2-year earlier \ros survey observation. The change in count rate 
(0.36 cts/s in the 1992 observation, 0.038 cts/s in 1990) is about a factor of 10.  

\section{Discussion}

\subsection{Comparison of different spectral models}

Based purely on the quality of the X-ray spectral fits, three different
models turn out to be successful: a black body, a flat powerlaw, and
a warm-absorbed powerlaw of canonical spectral index.
We favour the third one, since (besides hints for an absorption edge)
the first two are very unusual as judged from previous
X-ray observations of the class of Seyfert galaxies, and the presence
of a dusty warm absorber also fits to other properties and observations
of NGC 3786 (as detailed in Sect. 5.2). 

Thermal emission (although not necessarily of black-body-like shape)
or a flat powerlaw may arise from a nuclear starburst, 
i.e. hot gas and/or supernova remnants and X-ray binaries.
However, no other evidence of such a component is reported for NGC 3786, and  
the X-ray luminosity we find (a few 10$^{42}$ erg/s) overpredicts the one 
typically observed in galaxies with nuclear starbursts by a factor
of $\sim$10$^3$ (e.g. Fabbiano 1989). Furthermore, such emission
is not expected to vary by a factor of $\sim$10 within the timescale of 2 years.     
For a powerlaw emitted by the active nucleus, the one obtained (model 1) is {\em peculiarly flat},
with \G = --1.0, as compared to the canonical value of --1.9  
(e.g. Pounds et al. 1994, Walter et al. 1994, Nandra et al. 1997). 

\begin{figure}[thbp]
      \vbox{\psfig{figure=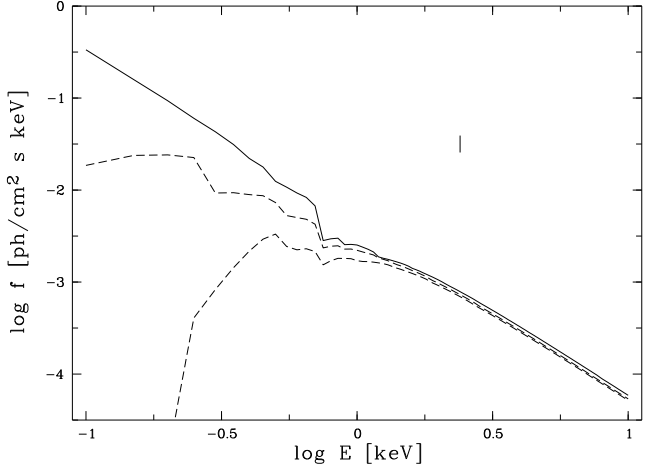,width=8.7cm,%
          bbllx=2.6cm,bblly=1.1cm,bburx=18.10cm,bbury=12.2cm,clip=}}\par
 \caption[wa_def]{Soft X-ray spectrum of NGC 3786 between 0.1 and 10 keV 
for several model descriptions.
Solid line: dust-free warm absorber (model 5, Table 1); 
dashed lines: dusty warm absorber, upper dashed line: model 7, lower dashed line: model 6. 
All spectra plotted have been corrected for cold absorption.
The vertical bar marks the upper end of the \ros energy range. }
 \label{wa_def}
\end{figure}

The presence of a warm absorber removes the need for invoking such an unusually 
flat powerlaw, and strong evidence for ionized absorbers has been found in
several of the well-studied Seyfert galaxies (cf. Fabian 1996 for a review).  
The amount of {\em cold} absorption present in NGC 3786 is more uncertain. 
A warm absorber model with Galactic \NH~ only already provides a successful fit.
The warm material is found to be relatively lowly ionized in this case,
$\log U \simeq -1.2$ (still somewhat higher than is typical for the 
broad line region (BLR) itself), and also contributes to the low-energy absorption
(Fig. \ref{wa_def}).    
A successful alternative description 
is a more highly ionized absorber combined with excess cold
absorption. 
The properties of the warm absorber are further discussed below in the context
of the Sy 1.8 classification of NGC 3786.    

\subsection{Sy 1.8 nature of NGC 3786} 

\subsubsection {Spectral properties} 

An interesting characteristic of intermediate type Seyferts like NGC 3786
is the often rather steep Balmer decrement 
of the BLR
as compared to Sy 1s. 
There are essentially two kinds of interpretations 
of the high observed H$\alpha_{\rm b}$/H$\beta_{\rm b}$ flux ratio 
offered in the literature:
One invokes reddening by dust along the line of sight, 
the other special properties of the BLR gas itself.   

In the first scenario, the intrinsic flux ratio is near 
the case-B recombination (Brocklehurst 1971) value, and the observed deviations
from that are caused by dust extinction (e.g. Osterbrock 1981).       
Some authors (e.g. Osterbrock 1981; Goodrich 1989, 1995) 
suggest the dust to be located in the outer BLR or just outside the
BLR.   
This ensures dust survival, since dust is expected to be destroyed in the bulk
of the BLR due to heating by the central radiation field  
and dust-gas interactions (e.g. Netzer 1990), 
and accounts for the usually less steep observed Balmer decrements of the NLR.
Maiolino \& Rieke (1995) propose the absorption 
to take place further outwards, in a 100 pc-scale torus (e.g. 
McLeod \& Rieke 1995) coplanar with the galactic 
disk (their Fig. 6).   

In the second interpretation (e.g. Canfield \& Puetter 1981; Rudy et al. 1988),
the flux ratio is {\em intrinsically} high.
High values can be reached in a narrow range of optical depths
and ionization parameters of the broad line gas (Canfield \& Puetter 1981).

Evidence for the dust interpretation in some objects was presented 
by Goodrich (1989, 1995) 
based on line variability studies. In another approach, Goodrich (1990) distinguished  
between both scenarios based on the Pa$\beta$ luminosity predicted in each case.   
For NGC 3786 he found no clear preference for either model. 

In the present study, we find strong evidence for excess absorption along the
line of sight to the X-ray continuum. 
However, due to the limited spectral resolution, the exact relative 
contributions of a warm and cold absorber cannot be clearly disentangled,
and several models that provide a successful fit 
are interesting in the light of the Sy 1.8 character of NGC 3786:   
model 5 (Table 1) is reminiscent of the Maiolino\&Rieke scenario,
if we interpret the cold absorption to represent the outer torus, 
and the high-$U$ (dust-free) warm absorber
the inner torus that is just grazed along the l.o.s.
On the other hand, 
in model 6 the dust mixed with the warm absorber would be located
between BLR and NLR, more similar to the Goodrich scenario.  

Given the different best-fit (warm or cold) absorbing columns 
we may ask which of the successful 
X-ray spectral models agrees best with the observed Balmer decrement.   
(We note that the optical and X-ray observations are not simultaneous, 
and time-dependent changes in the
amount of reddening cannot be excluded; see also next section). 
GO83 derive a flux ratio of the broad Balmer lines
of H$\alpha_{\rm b}$/H$\beta_{\rm b} \simeq$ 8.4, which corresponds to an extinction
of $E_{\rm B-V} \simeq$ 0.9 (assuming an intrinsic flux ratio of 3.1 and a Galactic
reddening law as in Osterbrock 1989). 
The narrow-line ratio, H$\alpha_{\rm n}$/H$\beta_{\rm n}$ = 2.5,
is near the case-B recombination value, indicating no reddening.
Assuming that dust is indeed responsible for the steep Balmer decrement
of the broad lines, 
and accompanied by an amount of cold gas as found in the Galaxy,
we expect an absorbing column of \NH~ $\simeq 5.5 \times 10^{21}$ cm$^{-2}$    
(e.g. Bohlin et al. 1978; see also Predehl \& Schmitt 1995). 
This is even larger than the value derived for the cold absorbing column in model 5 (Table 1). 
However, interestingly,
it is very near the {\em warm} column $N_{\rm w}$ obtained for the fit of a dusty warm absorber.  
Ionized material with internal dust between NLR and BLR would therefore offer a nice
explanation for the observed reddening properties. 

The dusty gas is expected to further reveal  
its presence in the  
optical--UV spectral region:    
{\em If} the observed X-ray and UV continuum originate in a region of comparable
extent as compared to that of the absorber, and travel along the same paths, 
the optical--UV continuum will be altered by gas absorption lines and 
dust reddening.   
The equivalent widths of the UV absorption lines CIV\,$\lambda$1549 and
NV\,$\lambda$1240 predicted to originate from the dusty warm absorber
(model 6; the values for models 5,7,4, respectively, are given in brackets)
 are $\log W_{\lambda}^{\rm CIV}$/$\lambda \simeq -2.94$ (--3.06, --3.01, --2.90) and 
$\log W_{\lambda}^{\rm NV}$/$\lambda \simeq -3.09$ (--3.13, --3.09, --2.98; 
for a velocity parameter 
$b$ = 60 km/s, Spitzer 1978).      
A search for such features in high-quality UV spectra will
certainly be worthwhile.  
The fraction of star light in the spectra of intermediate 
type Seyferts has been reported to be relatively high, which may indicate 
increased continuum reddening. For 
NGC 3786, the contribution is about 60\% (GO83). 
However, we note that the optical spectral index of the non-stellar powerlaw is 
$\alpha \simeq$ --1.0 (GO83), which does not strongly 
deviate from the mean slope found for a sample of Seyfert 1s ($\alpha \simeq$ --0.8;
Padovani \& Rafanelli 1988).   

The location of the dusty warm absorber is given 
by $r \simeq$ 0.055 pc/$\sqrt{n_{8}}$ (Sect. 3.2.2).
As noted in Sect. 1, Nelson (1996) found strong evidence for hot dust in the vicinity of
the active nucleus of NGC 3786. 
He estimates an extent of the dust cloud of 0.033 to 0.062 pc
from the nucleus.  
This component may be related to the warm absorber.   

No BLR reverberation mapping results exist for NGC 3786. 
Assuming a scaling of BLR radius with luminosity as found in 
other objects (e.g. Netzer 1990, Peterson 1993) results in a
smaller BLR radius in NGC 3786 than in NGC 3227. In the latter
object, the bulk of the BLR emission was found to arise at about 
0.01 pc (Salamanca et al. 1994, Winge et al. 1995). This value is
consistent with a location of the warm absorber outside the 
BLR in NGC 3786.    

As to the question whether the observed weak broad lines themselves may 
completely originate in
the warm material (instead of a `usual' BLR), we note that
the absorber-intrinsic luminosity in H$\beta$
is only $\approxlt$ 1/40 of the reddening-corrected observed
broad $L_{\rm H\beta}$ (for model 4, in which it is strongest; and much weaker
for the model involving dust mixed with the warm gas).

\subsubsection{X-ray variability} 

A factor of $\sim$10 variability in count rate is detected between the 2-year-separated
\ros observations.   
Pure {\em intrinsic} continuum variability is one possibility, 
another one is that    
we may have witnessed an event of increased
absorption during the \ros survey observation  
due to either (a) changes in the cold (or warm) {\em column density},  
or, (b) a reaction of the {\em ionization state} of the warm absorber to (small) changes in the
intrinsic luminosity. 

Events of changing emission-line extinction, on the timescale of years, have 
been reported by Goodrich (1989, 1995) for some intermediate Seyferts. 
The present observation may represent a similar event seen in X-rays. 

In a series of simulations we have determined the required change
in (a$_1$) \NH, or (a$_2$) $N_{\rm w}$, or (b) $L \Rightarrow U$   
to account for the observed drop
in count rate.  
First of all, we   
note that due to the relative `hardness' of the X-ray spectrum, 
comparatively high absorption will be required  
to change the count rate by a factor of 10 (since cold absorption mainly affects the
low-E region, which already only weakly contributes to the total count rate
during the pointed observation.)
The high expected columns are not unusual for BLR clouds or warm absorbers, though. 
More specifically, we find (numerical values are representatively given for  
 models 6 and 7 of Table 1): \\ 
(a$_1$) A factor of 16 change in the cold column density is necessary to
account for the low count rate of the survey observation, i.e. \NH$~\simeq~ 
1.9 \times 10^{22}$ cm$^{-2}$  
(model 7).  \\          
(b) In order to change sufficiently the absorption of the warm material 
by its response to a decrease in incident luminosity, we have to vary 
$L$ by a factor of $\sim$5 (model~7).
We consider this too large a factor to represent still 
an `elegant' solution (in the sense that one may then as well 
contribute all observed variability to changes in intrinsic luminosity). \\ 
(a$_2$) Finally, we varied $N_{\rm w}$.   
Since the warm material more effectively absorbs at higher energies within 
the \ros band than a cold absorber, we expect a smaller change in $N_{\rm w}$ to be 
required as compared to \NH. A value of $\log N_{\rm w} \simeq$ 22.45 
(model 6; corresponding to a factor of $\sim$5 increase) reproduces the observed 
count rate.  

Based on these findings we, again, favour the interpretation invoking 
the {\em dusty warm} absorber (model 6) which does not require
additional cold absorption in excess of the Galactic value,
i.e. accounts for all observations with the lowest number of 
components/parameters. In this model, then, a 
cloud of warm material may pass the line of sight
with a larger column along the l.o.s. during the survey observation.  
(Of course, a dust-free BLR cloud in motion (and  
constant properties of the dusty WA; model 7) would have had the same 
effect on the X-ray spectrum, although requiring a stronger change in \NH.  
Both models differ in the optical emission line reddening they predict. 
The reddening should change
in case of variable $N_{\rm w}$, but be unaffected in case of variable dust-free
cold \NH.       
Optical spectra taken simultaneously with the ROSAT observations   
would be valuable to discriminate between both possibilities.)   

Concerning observations with previous X-ray satellites, 
Persic et al. (1989) report an upper limit for the count rate 
in the direction of NGC 3786 obtained during the \heao survey 
of $R15 <$ 0.86 (with an error of 0.29),   
which converts to a luminosity of $L < 6.6 \times 10^{42}$ erg/s
in the energy range 2--10 keV.  
We predict (2--10 keV) luminosities of $L_1$ = 8.7 $\times 10^{42}$ erg/s
and $L_{4-7} \simeq$ 4 $\times 10^{42}$ erg/s
for models 1 and 4--7 of Table 1, respectively.  
This is consistent with the upper limit of Persic et al. (1989) within
the errors, i.e. with an intrinsically constant continuum.    
       
\mbox{             }  \\

\section{Summary and  conclusions}

We have reported the detection in X-rays of the Seyfert 1.8 galaxy
NGC 3786.
An analysis of the soft X-ray properties on the basis of \ros PSPC observations
was performed.  
The two most conspicuous features are (i)
evidence for strong excess absorption along the line of sight,
part (or all) of which may be caused by the presence of a warm 
absorber, and (ii) high-amplitude variability between the 2y-separated 
observations of a factor of $\sim$ 10 in count rate, with 
an intrinsic (0.1--2.4 keV) luminosity of 
\mbox{$L_{\rm x} \simeq$ a few $\times 10^{42}$ erg/s}
during the high-state.

These findings are discussed in the light of the Sy 1.8 nature of NGC 3786.
From several absorption models that provide a successful
description of the X-ray spectrum, we tentatively  
favour the one involving a warm absorber {\em with internal dust},
for which we obtain an ionization parameter of $\log U \simeq -0.8$,     
a column density of $\log N_{\rm w} \simeq 21.7$, 
a density-scaled distance from the nucleus of $r \simeq$ 0.055/$\sqrt{n_{8}}$ pc, 
and a density $n_{\rm H} \approxlt 10^8$ cm$^{-3}$.  
Placing the dusty material between BLR and NLR, this model is shown to 
provide an explanation 
for (a) the observed steepness of the broad Balmer decrement
(GO83) by dust reddening, 
and (b) the detected drop in the X-ray count rate by  
variability in the column density, e.g. by a cloud 
that crosses the line of sight.
The warm absorber may be related to the dusty torus and/or the
component of hot dust in NGC 3786 for which evidence was recently reported.   

NGC 3786 certainly is a very interesting object for further X-ray spectral
studies, particularly in combination with UV absorption line 
and optical reddening measurements. 

\begin{acknowledgements}
We thank Gary Ferland for providing {\em{Cloudy}}, 
and Hartmut Schulz for a critical reading of the manuscript. 
The \ros project is supported by the German Bundes\-mini\-ste\-rium
f\"ur Bildung, Wissenschaft, Forschung und Technologie (BMBF/DARA) and the Max-Planck-Society. 
This research has made use of the NASA/IPAC extragalactic database (NED)
which is operated by the Jet Propulsion Laboratory, Caltech,
under contract with the National Aeronautics and Space
Administration. 
The optical image shown is based on photographic data of the National Geographic Society -- Palomar
Observatory Sky Survey (NGS-POSS) obtained using the Oschin Telescope on
Palomar Mountain.  The NGS-POSS was funded by a grant from the National
Geographic Society to the California Institute of Technology.  The
plates were processed into the present compressed digital form with
their permission.  The Digitized Sky Survey was produced at the Space
Telescope Science Institute under US Government grant NAG W-2166.

\end{acknowledgements}

\end{document}